\documentclass[aps,prb,twocolumn,superscriptaddress,floatfix]{revtex4-2}
\pdfoutput=1 
\usepackage[colorlinks=true,citecolor=blue,linkcolor=blue,breaklinks=true]{hyperref}
\usepackage{amssymb} %math symbols
\usepackage{amsmath} %math typesetting
\usepackage{physics} %Provides macros for many physics symbols, including a convenient one for vectors
\usepackage{gensymb} %degree symbol = \degree in math or text mode
\usepackage{times} %Select Adobe Times Roman as default font. 
\usepackage{setspace} %Allows you to set spaces between lines. 
\usepackage{xfrac} %allows typeset fractions in the form `n/d' via the \sfrac command. Requires e-Tex, current LaTeX and Latex3 packages `template' and `xparse'.
\usepackage{placeins} 

\newcommand{\CTO}{CuTeO$_{4}$ }
\newcommand{\CTOns}{CuTeO$_{4}$}

\usepackage{array}
\newcolumntype{P}[1]{>{\centering\arraybackslash}p{#1}}

\begin{document}

% Use the \preprint command to place your local institutional report
% number in the upper righthand corner of the title page in preprint mode.
% Multiple \preprint commands are allowed.
% Use the 'preprintnumbers' class option to override journal defaults
% to display numbers if necessary
%\preprint{}

%Title of paper
\title{Quasi-1-dimensional Exchange Interactions and Short-range Magnetic Correlations in \CTO}

% repeat the \author .. \affiliation  etc. as needed
% \email, \thanks, \homepage, \altaffiliation all apply to the current
% author. Explanatory text should go in the []'s, actual e-mail
% address or url should go in the {}'s for \email and \homepage.
% Please use the appropriate macro foreach each type of information

% \affiliation command applies to all authors since the last
% \affiliation command. The \affiliation command should follow the
% other information
% \affiliation can be followed by \email, \homepage, \thanks as well.
%\email[]{Your e-mail address}
%\homepage[]{Your web page}
%\thanks{}
%\altaffiliation{}
\author{Zubia Hasan}
\thanks{These authors contributed equally to this work}
\affiliation{Institute for Quantum Matter, Department of Physics and Astronomy, The Johns Hopkins University, Baltimore, MD 21218, USA}
\author{Eli Zoghlin}
\thanks{These authors contributed equally to this work}
\affiliation{William H. Miller III Department of Physics and Astronomy, Johns Hopkins University, Baltimore, MD 21218, USA}
\author{Michal Winiarski}
\affiliation{Faculty of Applied Physics and Mathematics, Gdansk University of Technology, Narutowicza 11/12, 80‐
233 Gdansk, Poland}
\author{Kathryn E. Arpino}
\affiliation{Max Planck Institute for Chemical Physics of Solids, 01187 Dresden, Germany}
\author{Thomas Halloran}
\affiliation{Institute for Quantum Matter, Department of Physics and Astronomy, The Johns Hopkins University, Baltimore, MD 21218, USA}
\author{Thao T. Tran}
\affiliation{Department of Chemistry, Clemson University, Clemson, SC 29634, USA}
\author{Tyrel M. McQueen}
\email[]{mcqueen@jhu.edu.edu}
\affiliation{Institute for Quantum Matter, Department of Physics and Astronomy, The Johns Hopkins University, Baltimore, MD 21218, USA}
\affiliation{Department of Chemistry and Department of Materials Science and Engineering, The Johns Hopkins University, Baltimore, MD, 21218, USA}

%Collaboration name if desired (requires use of superscriptaddress
%option in \documentclass). \noaffiliation is required (may also be
%used with the \author command).
%\collaboration can be followed by \email, \homepage, \thanks as well.
%\collaboration{}
%\noaffiliation

\date{\today}

\begin{abstract}

\CTO has been proposed as a crystallographically distinct, yet electronic structure analog, of the superconducting cuprates. Here, we present detailed characterization of the physical properties of \CTO to address this proposal. Fitting of magnetic susceptibility data indicates unexpected quasi-1-dimensional, antiferromagnetic correlations at high-temperature, with a nearest-neighbor Heisenberg exchange of $J_{1}$ = 164(5) K. Low-temperature heat capacity measurements reveal a sizable $T$-linear contribution of $\gamma = 9.58(8) $ mJ $\text{mol}^{-1}$ $\text{K}^{-2}$, qualitatively consistent with expectations for a \textit{S} = $\frac{1}{2}$,  uniform, Heisenberg spin chain. Below $T \approx 40$ K the susceptibility shows an upturn inconsistent with quasi-1-dimensional behavior. While heat capacity measurements show no signs of magnetic order down to low-temperature, the upturn in the magnetic susceptibility coincides with the emergence of a diffuse peak (centered at $|\va*{Q}| \approx 0.7$ \r{A}) in the neutron diffraction data, indicative of peristent, short-range, antiferromagnetic order with a correlation length of $\xi$ = 10.1(9) \r{A} at $T =$ 10 K. The onset of non-linearity and hysteresis in the isothermal magnetization curves suggests the presence of a small ferromagnetic component. This persistent, short-range order is understood in the context of structural modeling of the X-ray and neutron diffraction data, which show the presence of a significant density of stacking faults. No evidence for substantive dopability is observed and \CTO appears, qualitatively, to have a larger band gap than predicted by density functional theory. We ascribe this finding to the inductive withdrawal effect from high-valence Te and suggest that superconductivity in copper tellurates is more likely to be found in compounds where there is a decreased reductive withdrawal effect from Te.

\end{abstract}

% insert suggested keywords - APS authors don't need to do this
%\keywords{}

%\maketitle must follow title, authors, abstract, and keywords
\maketitle

% body of paper here - Use proper section commands
% References should be done using the \cite, \ref, and \label commands
\section{Introduction \label{intro}}

High-temperature superconductivity in copper oxides is one of the most exciting emergent phenomena to result from strong electron-electron correlation effects \cite{dagotto1994,armitage2010,batlogg1991}. Significant progress has been made in the past 30 years in elucidating the key ingredients that underlie this behavior, with a layered crystallograpic structure, two-dimensional electronic structure, strong Cu-$d$ and O-$p$ orbital hybridization, and proximity to an insulating, magnetically ordered state, all believed to be important.

 \begin{figure*}
 \includegraphics[width =17.4 cm]{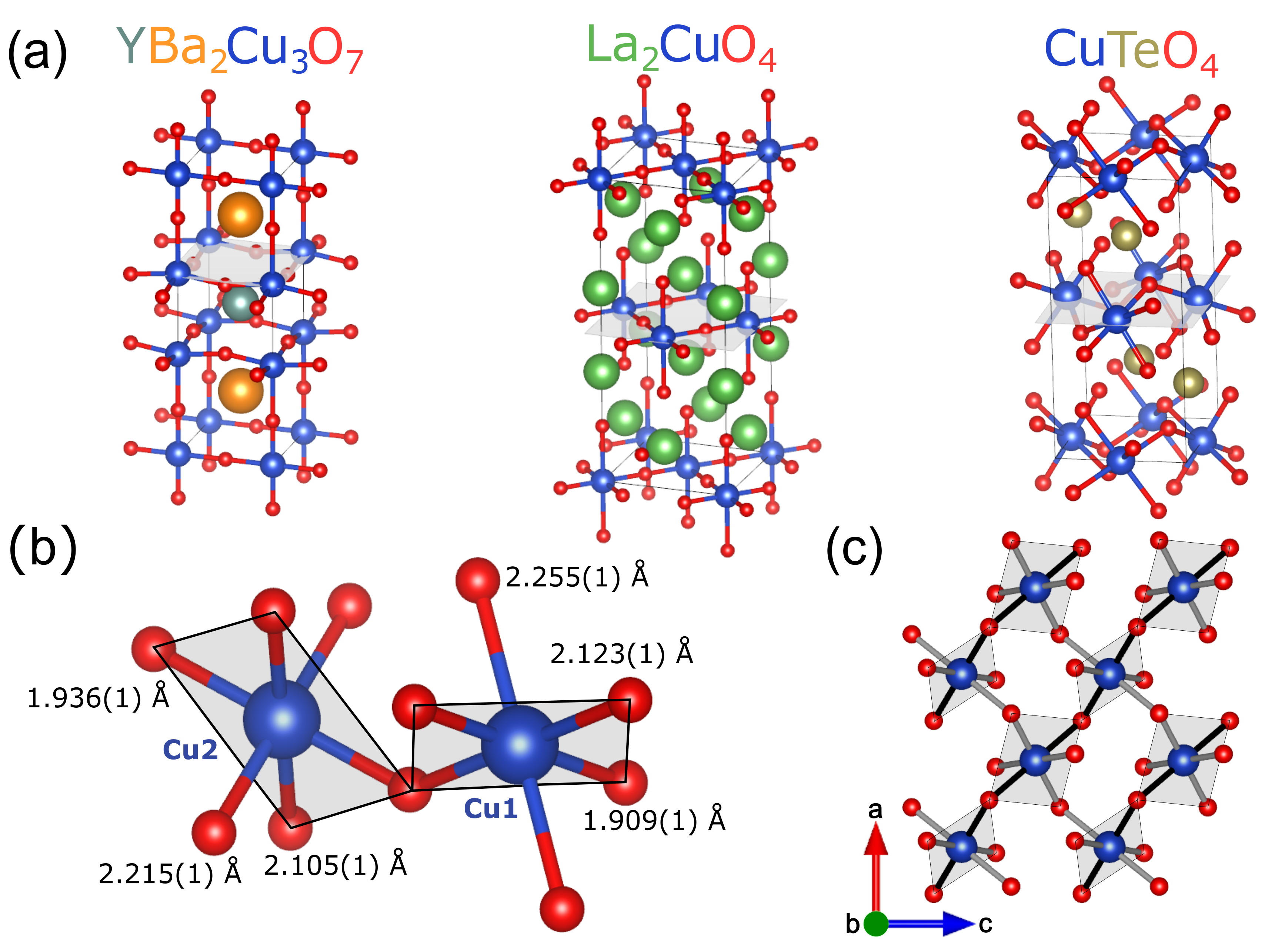}
 \caption{\label{fig:StructureComparison}\textbf{(a)} Comparison of the layered structures of known superconductors YBa$_{2}$Cu$_{3}$O$_{7}$ \cite{fernandes1991} and La$_{2}$CuO$_{4}$ \cite{rial1997} to \CTOns. The transparent grey planes indicate the Cu layers. All atomic visualizations were created with VESTA \cite{momma2011} and only Cu-O bonds are shown. \textbf{(b)} Illustration of the two varieties of distorted CuO$_{6}$ octahedra in \CTO  (Cu1 vs. Cu2) with the three unique bond distances labeled for each. The variation in coordination environment generates alternating twisted CuO$_{4}$ plaquettes in the Cu layers (grey transparent regions, corresponding to the two shorter Cu-O bonds in each octahedra). This strong buckling of the Cu layers is ultimately driven by shared oxygens with the TeO$_{6}$ octahedra. \textbf{(c)} Planar view of the Cu layer demonstrating how the corner-sharing connectivity of the alternating CuO$_{4}$ plaquettes creates a quasi-1D scenario, with the chains of plaquettes defined by the black Cu-O bonds. Note that the Cu-O-Cu pathways between different chains always involve one Cu-O bond that is not part of a CuO$_{4}$ plaquette.}
 \end{figure*}
 
Recent computational work identified \CTO as a previously unrecognized material containing these ingredients \cite{botana2017}. Figure \ref{fig:StructureComparison}(a) shows the reported crystal structure of \CTOns, alongside two other well-known cuprates, highlighting the layered nature. However, \CTO is distinct in that the Cu layer is built from distorted, highly angled, corner-sharing CuO$_{4}$ octahedra (Fig. \ref{fig:StructureComparison}(b)), with the Cu-O-Cu bond angle of 126.1$\degree$ closer to that of herbertsmithite \cite{shores2005}. Thus \CTO gives a possible new avenue to expand the known cuprate families and provides an opportunity to test which features are most important in ultimately producing superconductivity.

Despite being known for some time \cite{falck1978} there is scant experimental data on \CTOns, particularly in regard to its magnetic and electronic properties. This is likely due to
the difficulty in preparation, which was reported to require hydrothermal conditions at 650 \textdegree C and 2000 bar for two months in order to produce a multiphase mixture containing small crystals of \CTO as a side product. It is clear that development of a simpler, less onerous, route to preparing this material is needed to allow for detailed characterization of its properties and evaluation of its status as a potentially novel cuprate superconductor. 

Here, we report a novel method of synthesizing polycrystalline \CTO in mild hydrothermal conditions. The yellow color of the resulting powder suggests, qualitatively, a larger band gap than predicted by density functional theory (DFT) \cite{botana2017}, which we ascribe to the inductive withdrawal effect of hypervalent Te$^{6+}$ \cite{etourneau1992}. Carrier doping efforts with our synthesis method were not successful and the large band gap most likely rules out \CTO as a dopable host for high temperature superconductivity. These results suggest that future efforts at producing copper tellurate superconductors should look to minimize the inductive withdrawal effect by seeking materials with Te in a lower valence state. Nonetheless, enabled by our improved synthesis method, we present extensive characterization of the low temperature magnetic and electronic properties of \CTO through a combination of magnetic susceptibility, magnetization, and heat capacity, as well as X-ray and neutron scattering measurements.

At $T \geq$ 75 K our magnetic susceptibility data can be fit to a one-dimensional (1D) model corresponding to a uniform, $S$ = $\frac{1}{2}$, Heisenberg chain with an antiferromagnetic nearest-neighbor exchange. Such a finding is justified by a detailed examination of the crystal structure, which reveals how the distorted CuO$_{4}$ octahedra form quasi-1D chains. Elastic neutron scattering reveals the emergence of magnetic diffuse scattering below $T$ = 60 K. Since the AC magnetic susceptibility measurements are not consistent with a spin glass ground state we identify this diffuse scattering as due to the formation of short-range, AFM order. The observation of hysteresis in the isothermal magnetization curves, alongside the irreversibility seen in the zero-field cooled (ZFC) and field cooled (FC) DC magnetic susceptibilities ($T_{irr}$ = 40 K), suggests the presence of a ferromagnetic (FM) component, likely due to a canting of the AFM order.  Detailed analysis and modeling of our powder diffraction data indicates extensive faulting of the Cu layers. This has a clear impact on the magnetic properties, preventing the establishment of any long-range order down to $T$ = 0.1 K, as shown by our low-temperature heat capacity data, despite the onset of short-range correlations at much higher temperatures. A relatively large $T$-linear contribution to the heat capacity is observed at low temperatures ($\gamma$ = 9.58(8) mJ mole$^{-1}$ K$^{-2}$), qualitatively consistent with the expectation for a uniform, $S$ = $\frac{1}{2}$, AFM Heisenberg chain. 

\section{Methods}

\CTO was initially synthesized by Falck \textit{et al.} \cite{falck1978}, under high pressure and temperature. Here we produce polycrystalline \CTO via a modified method conducted at significantly lower pressure and temperature. A 1:3 molar ratio of CuO (2 mmol) and Te(OH)$_{6}$ (6 mmol) was used to make the reaction mixture. This mixture was placed in a 23 mL Teflon lined stainless steel autoclave with 10 mL of water and 0.03 mL of H$_{2}$O$_{2}$ \footnote{The inclusion of H$_{2}$O$_{2}$ seems to be crucial for the formation of \CTO in such mild conditions. Repetition of this experiment without H$_{2}$O$_{2}$ resulted in the formation of Cu$_{3}$TeO$_{6}$, the more thermodynamically favourable phase. The decomposition of \CTO into Cu$_{3}$TeO$_{6}$ alongside TeO$_{2}$ and O$_{2}$ at 510 \textdegree C is documented in \cite{gospodinov1992}. H$_{2}$O$_{2}$ quenches this decomposition by preventing the formation of TeO$_{2}$, oxidizing Te$^{4+}$ to Te$^{6+}$.}. A magnetic stir bar was added to the Teflon cup and the autoclave was closed and placed on a magnetic hot plate in a sandbath. A temperature of 210 \textdegree C was maintained for seven days with the help of a thermocouple placed next to the autoclave in the sand. A difference of a few degrees from the thermocouple sensing and the temperature within the autoclave is expected.  Various attempts were made to adapt the synthesis method for carrier doping (\textit{e.g.} substitution of Sb for Te) without success: no signs of dopant solubility in the parent phase were observed. We note here the observation that high-purity \CTO powder produced in this way is a bright yellow color.

The properties of the resulting polycrystalline powder of \CTO were characterized using a variety of methods. Phase purity was confirmed by powder x-ray diffraction data (PXRD) collected at room temperature using a Bruker D8 Focus diffractometer with a LynxEye detector and Cu K$\alpha$ radiation ($\lambda$ = 1.54 \r{A}$^{-1}$). Rietveld refinements were carried out using TOPAS version 4.2 \cite{coelho2018}. For magnetic properties measurements a small amount ($\approx$ 10 mg) of phase-pure \CTO powder was wrapped in a pouch of thin, polyethylene film and mounted inside a plastic straw. Care was taken to keep all mounting components clean to avoid contamination by extrinsic impurities (\textit{e.g.} Fe in dust from the laboratory environment) which could affect the magnetic measurements. DC magnetic susceptibility data ($\chi \equiv M/H$) were collected using the vibrating sample magnetometer (VSM) option of a Magnetic Properties Measurement System (MPMS, Quantum Design). AC magnetic susceptibility data were collected on the same instrument with a range of frequencies using an AC drive field of 3 Oe and a DC field of 100 Oe. Higher frequency data ($f$ = 10,000 Hz) were collected with the same field parameters using a Physical Properties Measurement System (PPMS, Quantum Design). The heat capacity of a pressed pellet was measured using the semiadiabatic pulse method in a PPMS equipped with a dilution refrigerator down to $T$ = 0.1 K. All data were collected on warming. The powder neutron diffraction (PND) data were obtained with the high-resolution diffractometer POWGEN at the Spallation Neutron Source at Oak Ridge National Laboratory. The sample was measured using frame 2 (center wavelength 1.5 \r{A}$^{-1}$) for a total of  6 $\times$ $10^{7}$ counts. Stacking faults in the material were simulated using DIFFaX version 1.813 \cite{treacy1991}. 

\section{Results}

\subsection{Crystal Structure} 

\begin{figure}
\includegraphics[width = 8.6 cm] {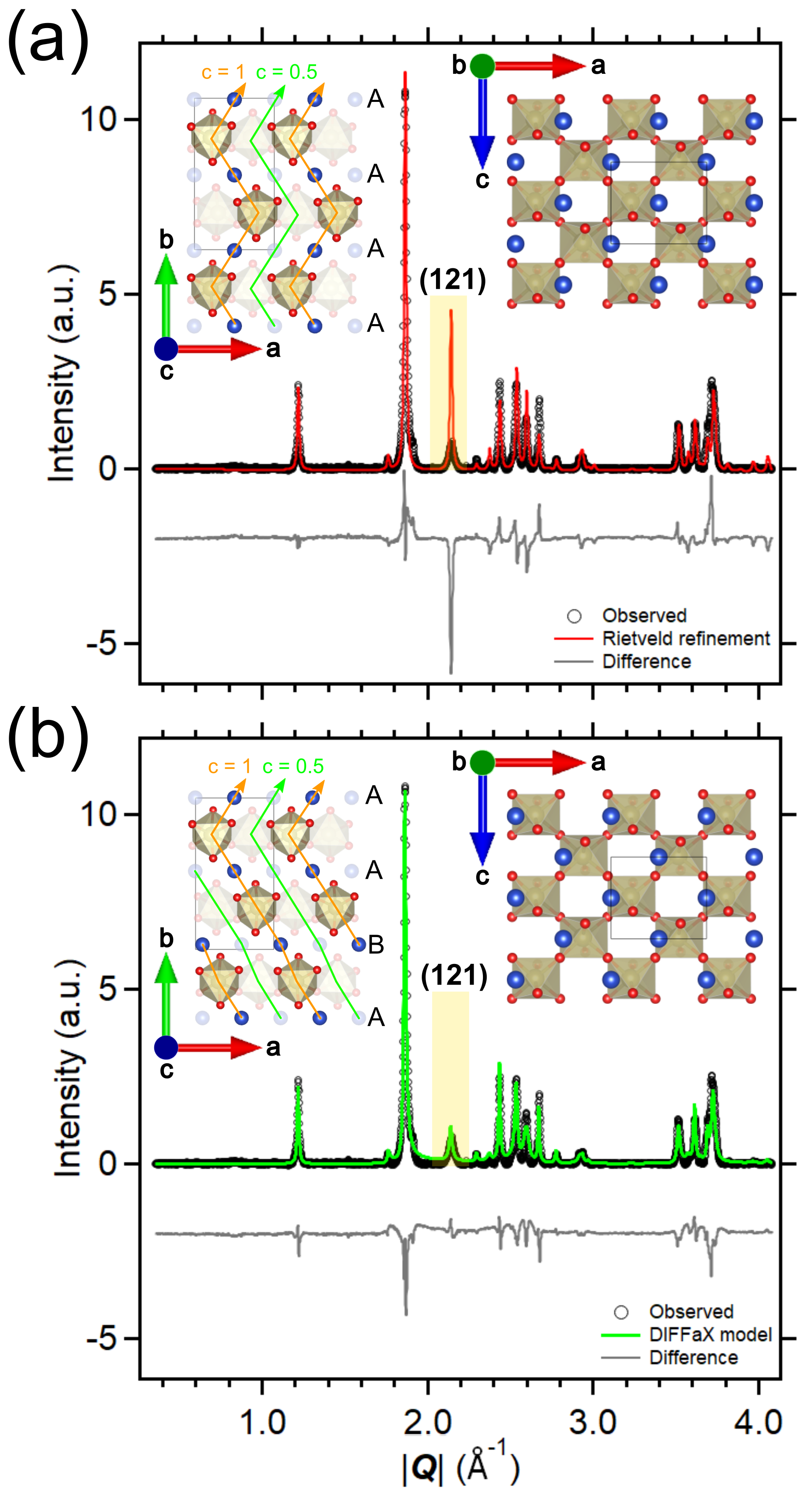}
\caption{\label{fig:refinement_diffax} \textbf{(a)} Rietveld refinement (red line) of the PXRD data ($\lambda$ = 1.54 \r{A}, black open circles), with difference curve below (grey line). The yellow shaded region highlights  the (121) Bragg peak. \textit{Inset}: The image on the left shows a side-on view of the stacking in the ideal, P21/\textit{n}, model. All of the Cu layers are the same leading to ``A'' stacking. The orange and green arrows show the interlayer connectivity of edge-sharing CuO$_{6}$ and TeO$_{6}$ octahedra, labeled by their distance along the \textit{c}-axis.  This distance is also represented by the difference in transparency. The O atoms which are only connected to Cu atoms have been omitted for clarity. The image on the right shows a top-down view of a single, un-faulted (``A''-type) Cu layer. \textbf{(b)} The model (solid green line) including stacking faults, as described in the text. The difference curve is shown only as a visual demonstration of the improvement in quality of the structural model. \textit{Inset}: Corresponding images as in (\textbf{a}) but with the addition of a faulted ``B'' layer, as described in the text. The image on the left shows how the addition of the fault modifies the interlayer connectivity. The image on the right shows a single, faulted (``B-type'') Cu layer.}
\end{figure}

Falck \textit{et al.} previously identified the crystal structure of \CTO as belonging to the P21/$n$ space group \cite{falck1978}. Rietveld refinement of this model to our PXRD data (Fig. \ref{fig:refinement_diffax}(a)) shows clear discrepancies, especially in the predicted intensity of the (121) Bragg peak. Such discrepancies indicate that this model is an incomplete description of the structure and suggest the influence of crystallographic disorder. The (121) is an allowed reflection specifically due to the Cu atoms on the P21/$n$ Wyckoff positions 2$a$ and 2$c$, which suggests disorder of the Cu layering as a potential origin. The presence of disordered layering (\textit{i.e.} stacking faults) is further supported by: (1) the inconsistent broadening of peaks in our experimental diffraction data, (2) the higher crystallographic symmetry suggested by $\beta$ $=$ 90$\degree$ in the monoclinic P21/$n$ model coupled with the fact that the structure could not be solved in a higher symmetry spacegroup, and (3) the statement from Falck \textit{et al.} that their P21/$n$ model could also be described as the TeO$_{4}$ layers belonging to spacegroup P$nma$, along with the lower-symmetry Cu atoms possessing only $n$-glide symmetry \cite{falck1978}. 

Based on the ideal structural model we considered the possibility that a layer of Cu atoms might occur on P21/$n$ Wyckoff positions 2$b$ and 2$d$ instead of 2$a$ and 2$c$ (see insets of Figs. \ref{fig:refinement_diffax}(a) and \ref{fig:refinement_diffax}(b)). While the two are similar, Cu atoms in the faulted layer have a slightly greater overlap with the TeO$_{6}$ octahedra of the underlying layer. As both original and faulted Cu layers have the same intralayer arrangement the connectivity of the Cu atoms within each layer remains the same. To test this possibility we used DIFFaX to simulate structural models with 0$\%$ \textendash 100$\%$ faulted Cu layers. Inspection of these models indicated that a probability of 70$\%$ original and 30$\%$ faulted Cu layers (Fig. \ref{fig:refinement_diffax}(b)) significantly improves qualitative agreement with the experimental data in terms of the predicted peak intensities and widths. Note that this model was \textit{not} arrived at via a least-squares minimization routine so the reported percentage of faulted layers is not truly quantitative. Applying this faulted model to our powder neutron scattering data (not shown) also resulted in a significant improvement in qualitative agreement relative to a Rietveld refinement with the ideal structure.

\subsection{Magnetism}

 \begin{figure}
 \includegraphics[width = 8.4 cm] {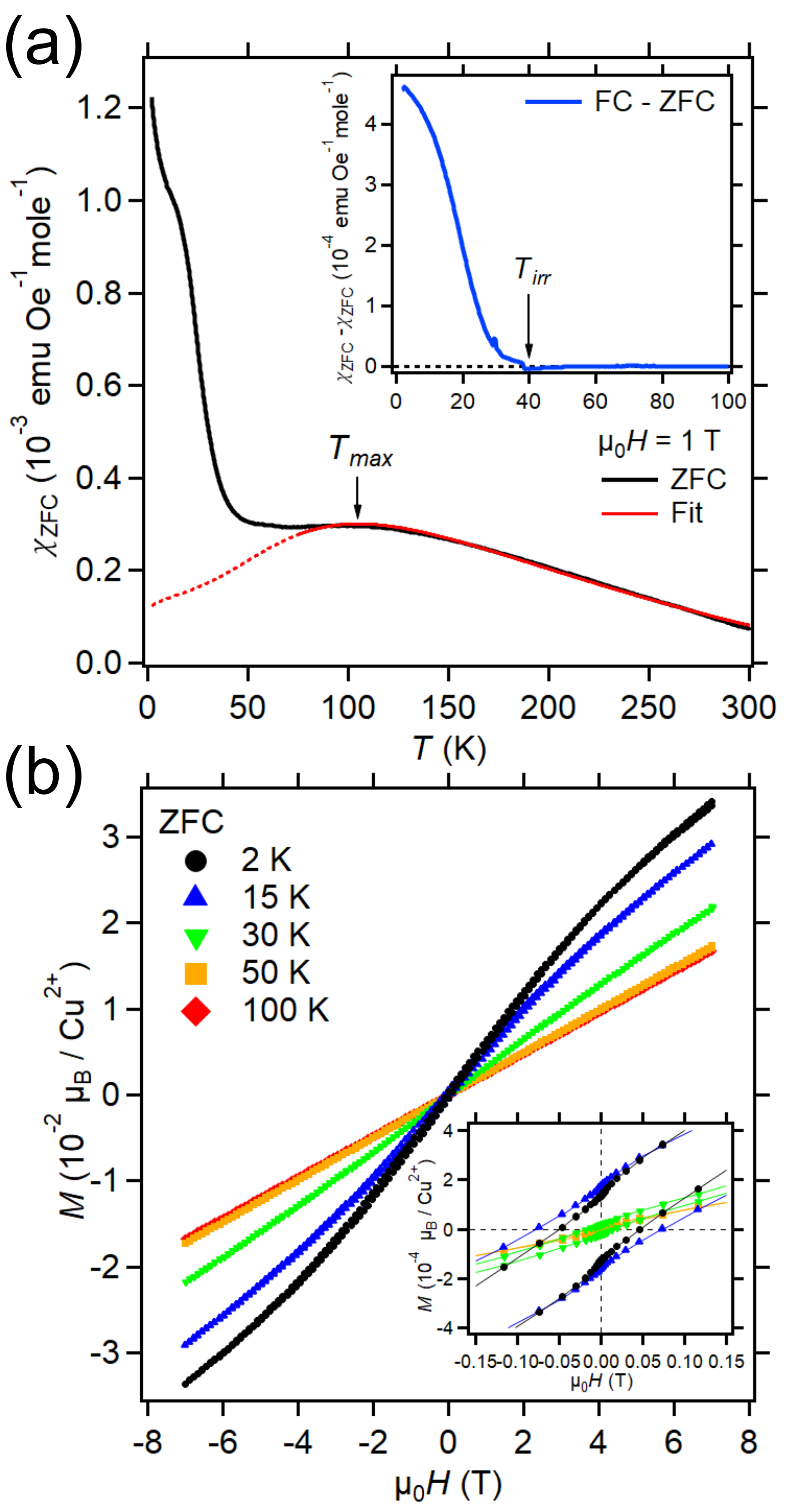}
 \caption{\label{fig:DC_magChar}\textbf{(a)} ZFC magnetic susceptibility. The fit (solid red line) uses the AFM alternating exchange model described in the text; the dashed red line represents extrapolation of the fit function to lower temperature. The black arrow indicates the location of the weak maximum at $T_{max}$ = 100 K. \textit{Inset}: Plot of the FC susceptibility, $\chi_{FC}$, minus the ZFC susceptibity, $\chi_{ZFC}$. The black arrow indicates the onset of irreversibility, $T_{irr}$ = 40 K, as determined by the bifurcation of the FC and ZFC curves. All data were collected using $\mu_{0}H$ = 1 T. \textbf{(b)} Isothermal magnetiation curves collected at various temperatures for fields up to $\mu_{0}H$ = \textpm7 T. \textit{Inset}: Closer view of the low-field region. Solid lines are a guide to the eye.}
 \end{figure}

Temperature dependent, ZFC, DC magnetic susceptibility data for \CTO are displayed in Fig. \ref{fig:DC_magChar}(a). No clear signature of three-dimensional, long-range ordering is observed down to $T$ = 2 K. Furthermore, there is no region in the susceptibility data that could be accurately fit with the Curie-Weiss model; such behavior is typical of quasi-1D systems \cite{Motoyama1996,Urushihara2020, khuntia2010}. A weak maximum,  another characteristic of quasi-1D systems \cite{bonner1964}, is observed at $T_{max}$ = 100 K, and we find that for $T \geq$ 75 K, the data can be reasonably well fit by the expression:

\begin{equation}
\chi(T)=A \chi^*\left(\alpha, \frac{k_{B}T}{J_{1 }}\right)+\chi_0 ,
\end{equation}where $A = \frac{N_{A}g^{2}\mu_{B}^{2}}{J_{1}k_{B}}$ is an overall scaling factor and $\chi_{0}$ is a temperature independent contribution. $\chi^*(\alpha, \frac{T}{J_{1}})$  is a fitting function derived from a high-temperature series expansion for the susceptibility of the AFM alternating exchange model:

\begin{equation}
\chi^*(\alpha, t= k_{B}T/J_{1})=\frac{\mathrm{e}^{-\Delta^*(\alpha) / t}}{4 t} \mathcal{P}_{\mathrm{m}(8)}^{(7)}(\alpha, t),
\end{equation}where $\Delta^*(\alpha) = 1 - \frac{1}{2}\alpha - 2\alpha^{2} + \frac{3}{2}\alpha^{3}$ and $\mathcal{P}_{\mathrm{m}(8)}^{(7)}(\alpha, t)$ is a modified Pad{\'e} approximant, as described in Johnston \textit{et al.} \cite{johnston2000}.

 The AFM alternating exchange model describes a Heisenberg chain with two nearest-neighbor, AFM, intrachain exchange interactions, $J_{1}$ and $J_{2}$, with $J_{2} \leq J_{1}$  and $\alpha = J_{2}$ / $J_{1}$. Note that we follow the convention employed by Johnston \textit{et al.} in which a positive sign is used for the AFM exchange constants. At $\alpha$ = 0, the model recovers the isolated spin dimer model while $\alpha$ = 1 recovers the uniform AFM Heisenberg chain model. The results from the best fit are given in Table \ref{tab:susceptFit}. Our fitting indicates that, for $T  \geq$ 75 K, \CTO behaves as system of spin chains with uniform ($\alpha$ = 1.000(1)) nearest-neighbor exchange interactions. This is suprising in the context of previous work, which considered the Cu layers as effectively 2D. We return to this point in the discussion. Based on the value of the scale factor, $A$, we extract $g$ = 1.4(2), which is significantly lower than the expected value of 2.2. However, $A$ is more strongly influenced by extrinsic factors (\textit{e.g.} the error in the mass of the sample) than other fit results, meaning the systematic error is likely larger than what we have quoted, precluding any definitive interpretation. We note here that the magnitude of $\chi_{0}$ is somewhat larger than the value expected for the core diamagnetism based on ideal ion valences ($\approx -7 \times 10^{-5}$ emu Oe$^{-1}$ mole$^{-1}$ \cite{bain2008}). This is expected since the temperature-independent, diamagnetic contribution from the sample holder, which has not been removed, is comparable to the total raw signal from the sample and therefore has a strong influence on the fitted value of $\chi_{0}$.
 
 \begin{table}
\caption{\label{tab:susceptFit}  Results from the fit to $\chi_{ZFC}$ ($T \geq$ 75 K) with the AFM alternating exchange model described in the text and shown in Fig. \ref{fig:DC_magChar}(a). Errors were obtained primarily by varying the range of the fit.}
\begin{ruledtabular}
\begin{tabular}{P{4.29cm}|P{4.29cm}}
Parameter & Fit result \\ \hline
& \\
$A$ (emu Oe$^-1$ mole$^-1$) & 4.7(2) $\times$ $10^{-3}$ \\
& \\
$J_{1}$ (K) &  164(5) \\
& \\
$\alpha$ &  1.000(1) \\
& \\
$\chi_{0}$ (emu Oe$^{-1}$ mole$^{-1}$) & -3.9(3) $\times$ $10^{-4}$
\end{tabular}
\end{ruledtabular}
\end{table}
 
Below the temperature regime where $\chi$($T$) can be well fit by this 1D model, an upturn in the susceptibility is visible near 40 K. To probe the character of this feature, field-cooled DC magnetic susceptibility data were taken and the subtraction ($\chi_{FC} - \chi_{ZFC}$) is shown in the inset of Fig.  \ref{fig:DC_magChar}(a). These data show a clear onset of irreversibility at $T_{irr}$ = 40 K, coincident with the upturn seen in the $\chi_{ZFC}$. Further magnetic characterization was carried out by collecting a series of isothermal magnetization curves ($M$ versus $H$), which are shown in Fig. \ref{fig:DC_magChar}(b). The magnetization curves are linear across the measured field range at higher temperatures, but begin to change slope and develop a slight curvature and small hysteresis below $T$ = 50 K. A remnant magnetization of 1.3 $\times 10^{-4} \mu_{B}$ / Cu$^{2+}$ is observed at 2 K.
 
  \begin{figure}
 \includegraphics[width = 8.6 cm]{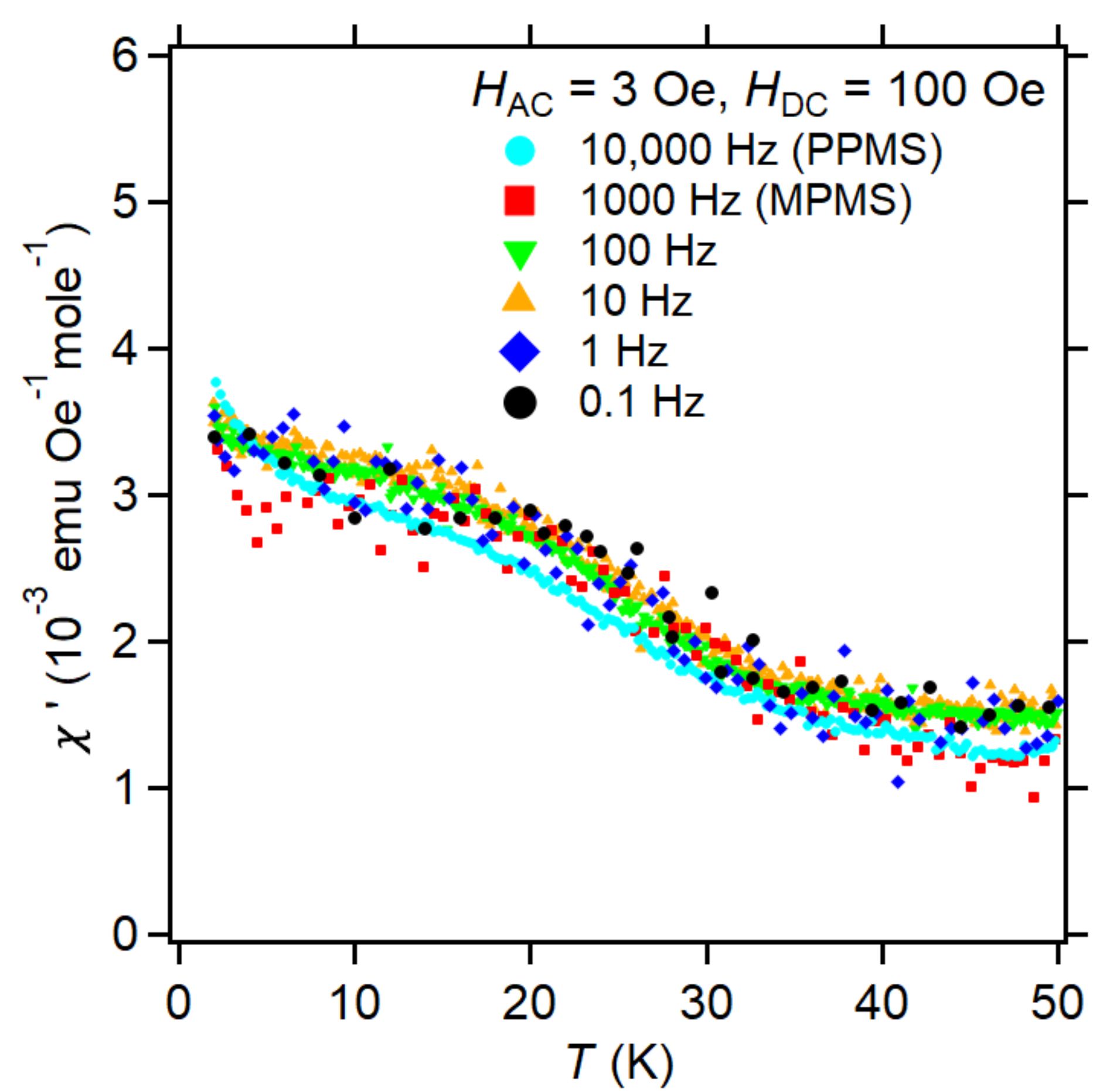}
 \caption{\label{fig:ACsuscept}AC magnetic susceptibility data collected with a range of AC drive frequencies ($f$ = 0.1 Hz - 10,000 Hz) using constant magnetic field values}. 
 \end{figure}

As a complement to the DC data, AC magnetic susceptibility data were collected across six orders of magnitude in frequency (Fig. \ref{fig:ACsuscept}). The data show a small upturn below about 35 K, consistent with the significant change in slope of the magnetization curves observed below that temperature (see Fig. \ref{fig:DC_magChar}(b)). Within the noise limitations of our measurements this feature does not appear to shift appreciably with frequency.
 
  \begin{figure}
 \includegraphics[width = 8.6 cm] {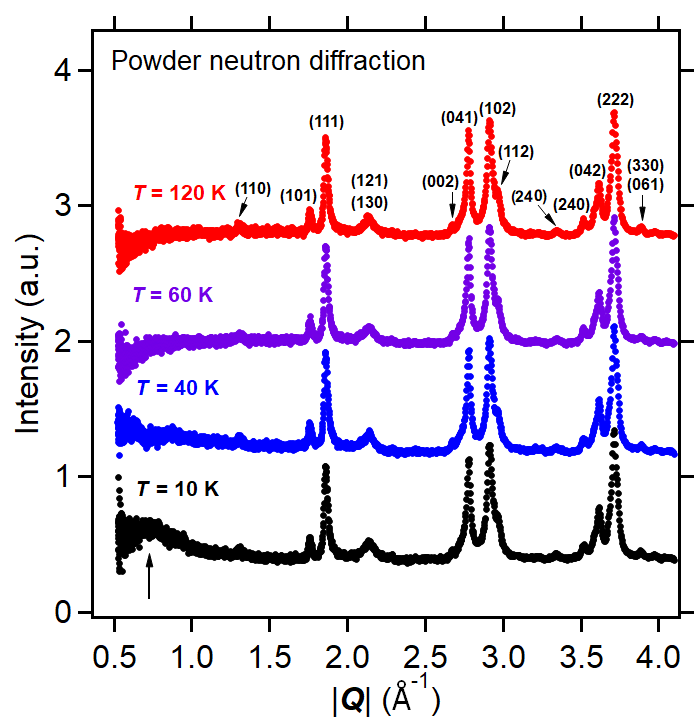}
 \caption{\label{fig:POWGEN_Tdep} Elastic neutron scattering data at various temperatures, offset for clarity. The broad, $T$-dependent, feature at low-$|\va*{Q}|$ is indicated by the black arrow at the bottom of the figure. The (nuclear) peaks corresponding to \CTO are labeled by their ($hkl$) value.}
 \end{figure}
 
\begin{figure}
\includegraphics[width = 8.6 cm] {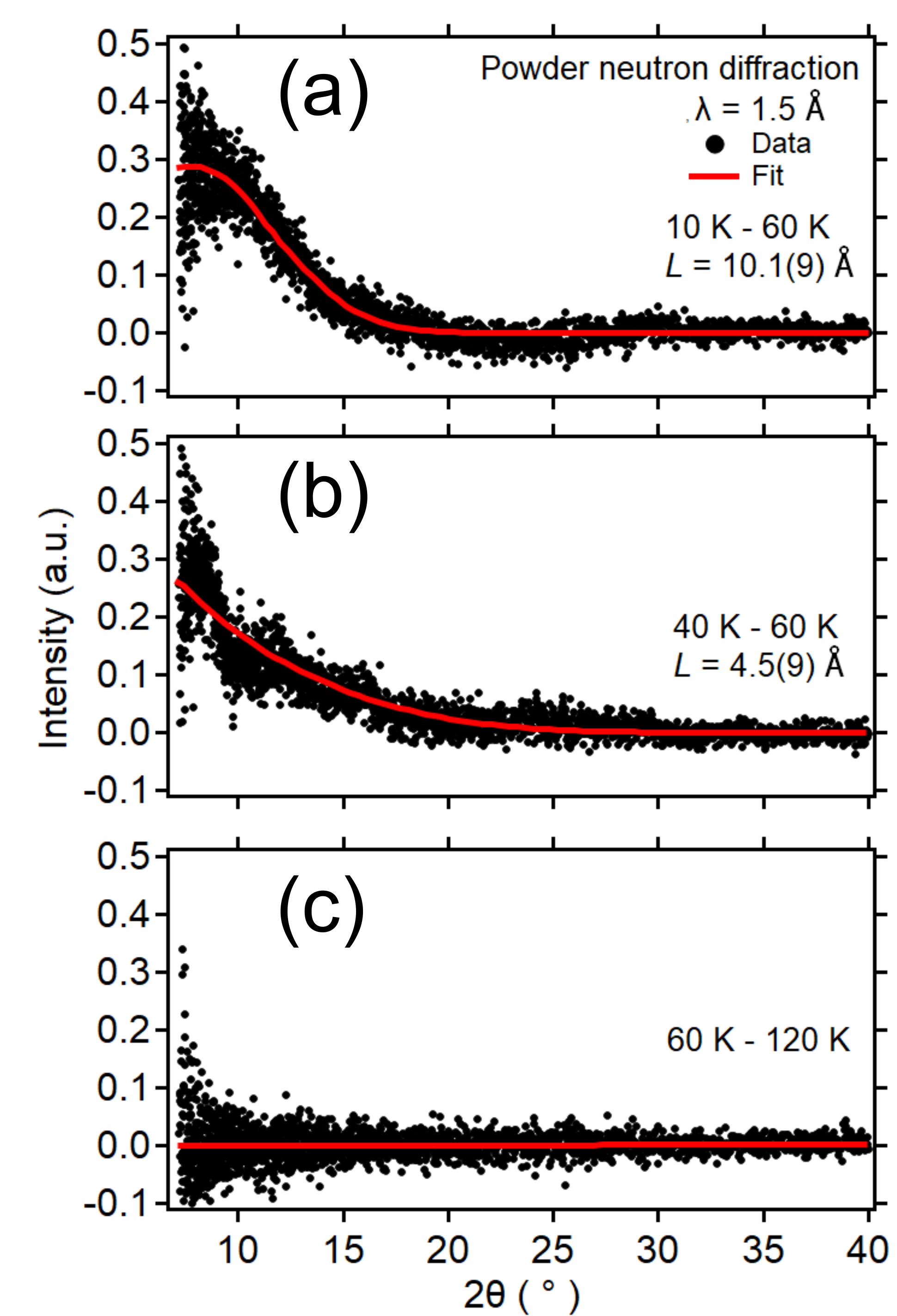}
\caption{\label{fig:warrenline shape} High-temperature subtracted elastic neutron scattering data (filled black circles) at various temperatures. The conversion to scattering angle was performed using the center wavelength ($\lambda$ = 1.5 \r{A}) for the frame 2 wavelength distribution at POWGEN. The red line is a fit to the data using the Warren line shape, as described in the text. \textbf{(a)} $T$ = 10  - 60, \textbf{(b)} 40K - 60, and \textbf{(c)} 60  - 120 K.}
\end{figure}

To further elucidate the nature of the low-temperature magnetism, time-of-flight (TOF) elastic neutron scattering was carried out at $T$ = 10 , 40 , 60 and 120 K (Fig. \ref{fig:POWGEN_Tdep}). All Bragg peaks could be indexed using the reported crystallograpic structure and no additional magnetic Bragg peaks characteristic of long-range order were observed at lower temperature. However, a broad, diffuse peak emerges at low $|\va*{Q}|$ ($\approx$ 0.7 \r{A}$^{-1}$) as the temperature is decreased below $T$ = 60 K. For $T \geq$ 60 K, the scatttering intensities remain essentially constant. This is most clearly shown by subtracting the high-temperature data from the low-temperature data, as is done in Fig. \ref{fig:warrenline shape}. The width of this diffuse peak is well-described by a Warren line shape \cite{warren1941}, which is characteristic of short-range order along the corresponding crystallographic direction. The Warren line shape function is: 

\begin{equation}
   P(2\theta) = C \frac{1+ \cos^{2} 2\theta }{2 (\sin \theta)^{\frac{3}{2}}} \frac{\sqrt{\pi}}{2e^{a^{2}}} ,  
\end{equation} where $ C = KmF_{hkl}^2$($\frac{\xi}{\lambda \sqrt{\pi}})^{1/2}$  and $a = \frac{2\xi\sqrt{\pi}}{\lambda}(\sin(\theta) - \sin(\theta_{o}))$. Here $K$ is an overall scale factor and $F_{hkl}$ is the structure factor for reflection ($hkl$) centered at the scattering angle of $2\theta_{o}$, with multiplicity $m$ and average correlation length $\xi$. Figure \ref{fig:warrenline shape} shows the data from Fig. \ref{fig:POWGEN_Tdep} as a function of scattering angle ($2\theta$) with a high-temperature subtraction applied. Fitting to the Warren line shape, the extracted correlation length is $\xi$ = 4.55(6) \r{A} at 40 K, rising to $\xi$ = 10.1(9) \r{A} at 10 K (Table \ref{tab:warren}). The value of $\theta_{o}$ for $T$ = 10 K is equivalent to a $|\va*{Q}|$-value of 0.74 \r{A}, in agreement with the experimental data in Figure \ref{fig:POWGEN_Tdep}. For the $T$ = 40 K fit, $\theta_{o}$ was constrained to the $T$ = 10 K value. 

\begin{table}
\caption{\label{tab:warren}  Fit parameters from the Warren line shape analysis. Values in italics were held constant during the fitting process.}
\begin{ruledtabular}
\begin{tabular}{ P{2.15cm} |P{2.15cm} |P{2.15cm} |P{2.15cm} }
    $T$ (K) & $C$ & $\theta_{o}$ (\textdegree) & $\xi$ (\r{A})\\ \hline
    & & &  \\  
     10  & 0.732(5) & 5.07(1) & 10.1(9) \\
     & & &  \\   
     40 & 0.505(3) & \textit{5.07} & 4.5(9) 
\end{tabular}
\end{ruledtabular}
\end{table}

\subsection{Heat Capacity}

\begin{figure}
\includegraphics[width = 8.6 cm] {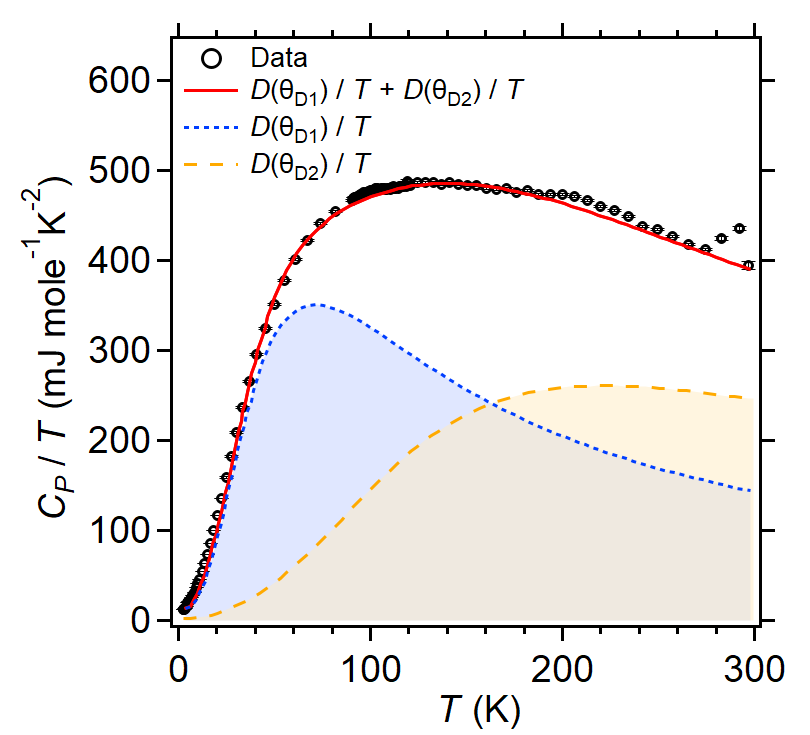}
\caption{\label{fig:heatCapacity_highT} Temperature dependence of the heat capacity of polycrystalline \CTO in the range $T$ = 2 -- 300 K. The slight hump near 200 K and the peak near 300 K are transitions of the Apiezon N grease used to affix the sample \cite{kreitman1971}. The total fit function is shown with the solid red line while the dashed blue and orange lines and accompanying shading show the individual Debye terms.}
\end{figure}

Figure \ref{fig:heatCapacity_highT} shows heat capacity data down to $T$ = 2 K. No sharp features associated with phase transitions are seen in this temperature range.  Initial attempts to describe the data with a single Debye or Einstein term did not yield a satisfactory fit. However the data were well fit for $T \ge$ 4 K by the sum of two Debye terms: 
\begin{equation}
    \frac{C_{Debye}}{T} = \frac{D(\Theta_{D1},T)}{T} + \frac{D(\Theta_{D2},T)}{T}.
\end{equation}

$D(\Theta,T)$ is the Debye expression for the heat capacity:
\begin{equation}
D(\Theta_{D}, T) = 9s_{D}R\left(\frac{T}{\Theta_{\mathrm{D}}}\right)^3 \int_0^{\Theta_{\mathrm{D}} / T} \frac{x^4 e^x}{\left(e^x-1\right)^2} d x,
\end{equation} where $\Theta_{D}$ is the Debye temperature and $s_{D}$ is the number of oscillators per formula unit (the ``oscillator strength''). The fit function was constrained such that $C_{Debye} \le C_{P}$($T$) at higher temperatures in order to prevent overestimation of the phonon contribution to the experimentally measured value. Fitting using the combination of a Debye and an Einstein term was also performed, but this gave worse agreement than the sum of two Debye terms.

The physical interpretation for the utilized model is as follows. Given the presence of multiple atoms in the unit cell with a large variation in atomic mass the optical modes of the phonon spectrum will become well-separated from the acoustic modes in energy. This can effectively create two seperate subsystems leading to two terms in the heat capacity with different characteristic temperatures. If the optical modes are truly localized their contribution to the heat capacity can be captured by an Einstein term with the rest of the lattice contribution captured by a Debye term. However, if the optical modes possess significant dispersion than their contribution to the heat capacity can take a Debye-like character. Based on our fitting attempts this appears to be the case for \CTOns. Additional motivation for the two Debye term model is based on the presence of significant disorder -- the stacking faults clearly inidcated by the diffraction data -- which should smear out the phonon dispersion curves in energy, pushing the optical modes further from an Einstein-like character. 

The results of the two Debye term fit are shown in Table \ref{tab:debyeModel}; we note that the oscillator strengths sum to 5.86, close to the value of six atoms per formula unit for \CTOns. This agreement, which was not obtained with other models, lends creedence to the use of two Debye terms in fitting the data. Within this temperature range the expected Dulong-Petit value of 3$R$ is not recovered. This can be explained by the high value of $\Theta_{D2}$ since, within the Debye model, the plateau to the Dulong-Petit value occurs well above the Debye temperature. 

\begin{figure}
\includegraphics[width = 8.6 cm] {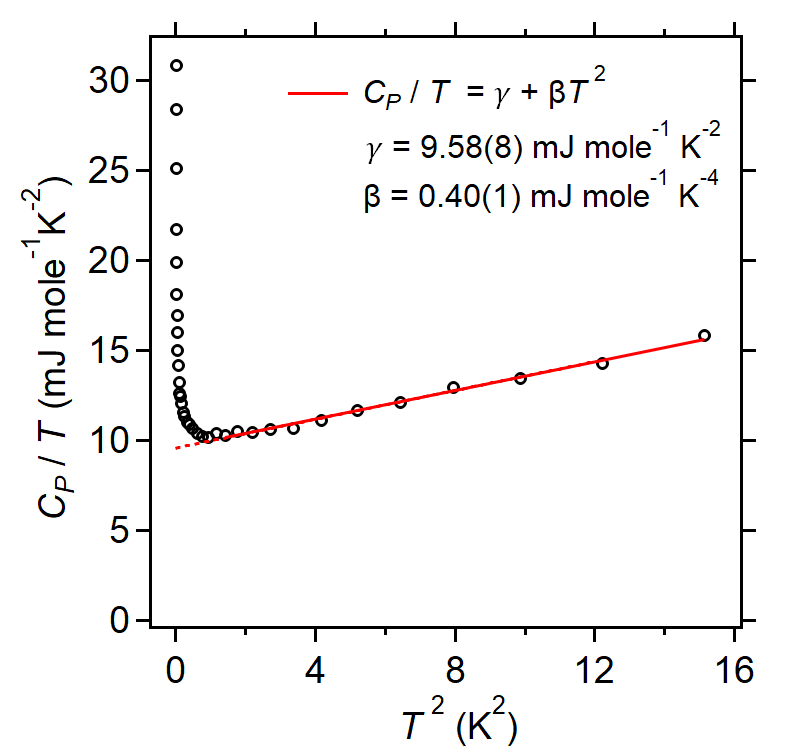}
\caption{\label{fig:heatCapacity_lowT} Heat capacity data (open circles) for the range $T$ = 0.1 - 4 K. The fit (solid red line) was performed in the temperature range where the data appears linear ($T \geq$ 1.2  K). The dashed red line is an extrapolation of the fit function to $T$ = 0 K.  }
\end{figure}

The measured heat capacity in the range $T$ = 0.1 --  4 K is shown in Fig. \ref{fig:heatCapacity_lowT}. Fitting the linear region of these data to the expression: $\frac{C_p}{T} = \beta T^2 + \gamma$, we obtain $\gamma = 9.58(8) $ mJ $\text{mol}^{-1}$ $\text{K}^{-2}$ and $\beta = 0.40(1)$ mJ $\text{mol}^{-1}$ $\text{K}^{-4}$. Employing the low-temperature limit of the Debye model, this value of $\beta$ yields $\Theta_{D}$ = 307(3) K, closer to $\Theta_{D1}$ than $\Theta_{D2}$. This agrees with our Debye analysis of the higher temperature data since $D(\Theta_{D1},T)$ has the larger contribution to the total heat capacity at lower temperatures (see Fig. \ref{fig:heatCapacity_highT}).

\begin{table}
\caption{\label{tab:debyeModel}Fit results from modeling the high temperature heat capacity data as shown in Fig. \ref{fig:heatCapacity_highT}.} 
\begin{ruledtabular}
\begin{tabular}{ P{4.29cm} | P{4.29cm} }
    Parameter & Value \\ \hline
    & \\
    $\Theta_{D1}$ & 257(1) K\\ 
    & \\
    $s_{D1}$ & 1.78(2) \\ 
    & \\
    $\Theta_{D2}$ & 794(13) K\\ 
    & \\
    $s_{D2}$ & 4.08(6) \\
\end{tabular}
\end{ruledtabular}
\end{table}

\section{Discussion} 

This study was motivated by the proposal of \CTO as a candidate high-$T_c$ copper tellurate based on apparent similarities with the parent phases of superconducting cuprates, including a layered crystal structure and computationally predicted two-dimensional electronic structure and antiferromagnetic, insulating ground state \cite{botana2017}. Our experimental results challenge this picture in several ways. 

The characterization of \CTO as a \textit{dopable} parent phase for superconductivity is based on the small band gap of 0.13 eV calculated \textit{via} DFT in Botana \textit{et al.} \cite{botana2017}. However this is qualitatively inconsistent with the observed yellow color of the polycrystalline powder mentioned previously. The color of \CTO is similar to that of CdS -- which has a direct band gap of 2.42 eV \cite{lincot2006} --  suggesting a band gap on the order of a few eV. While the GGA functional used for the calculations is known to underestimate the band gap, closer inspection of the band structures suggests that the origin of the underestimation is not solely due to the exchange correlation functional used. We propose that the large underestimation may be due, in part, to the failure of DFT to consider the effect of hypervalent cations such as Te$^{6+}$, which have a strong inductive withdrawal effect on O$^{2-}$ \cite{etourneau1992}. The resulting covalent character of the Te-O  bond likely also explains the inability to synthesize carrier doped samples \textit{e.g.} CuTe$_{1-x}$Sb$_{x}$O$_{4}$. 

A further challenge is that, as demonstrated by our fitting of the magnetic susceptibility (Fig. \ref{fig:DC_magChar}(a)), the magnetism of \CTO above $T$ = 75 K is well-described by a uniform quasi-1D, AFM Heisenberg spin chain model. While the presence of extensive faulting of the Cu layers confirms the layered nature of the crystallographic structure, a detailed examination of the connectivity of the CuO$_{4}$ plaquettes clarifies how quasi-1D magnetic interactions can arise. 

The Cu layers are composed of two types of distorted CuO$_{6}$ octahedra with distinct Cu-O bond length hierarchies. Figure \ref{fig:StructureComparison}(b) shows the basic unit of two corner-connected octahedra, where the relative rotation (\textit{i.e.} buckling of the layer) is driven by the edge-sharing connection to neighboring TeO$_{6}$ octahedra. The CuO$_{4}$ plaquettes are defined by the shortest two Cu-O bonds in each octahedra and, due to the rotations, are only corner-connected to their neighbors along the [101] direction. The dominant superexchange interactions are then along this Cu-O-Cu pathway (Fig. \ref{fig:StructureComparison}(c), black line), mediated by the  d$x^{2}-y^{2}$ orbitals lying within the plane of individual CuO$_{4}$  plaquettes. This situation is similar to that of Na$_{2}$Cu$_{2}$TeO$_{6}$, where quasi-1D behavior is observed despite a layered, 2D, crystallographic structure; however, in that case, the intrachain exchange interactions were found to alternate strongly ($\alpha$ = 0.10(1)) indicating chains of weakly interacting dimers \cite{xu2005}. A contrasting example is that of the monoclinic polymorph of SeCuO$_{3}$, where the dimensionality is further reduced, such that the magnetic susceptibility is best described by a model of isolated Cu tetramers \cite{kivkovic2012}. Comparison of these compounds highlights how the dimensionality of magnetic phase behaviors in cuprates depends strongly on the details of the connectivity of the lattice and the resulting exchange interactions.

 Our result of $J_{1}$ = 164(5) K is roughly half of that expected from the 126.1$\degree$ Cu-O-Cu bond angle, which theory indicates should correspond to $J_{1} \approx 400$ K \cite{rocquefelte2012}. This likely reflects the greater ionicity of the Cu-O bonds in \CTO due to the large inductive withdrawal effect of neighboring Te$^{6+}$ ions \cite{etourneau1992}. The observed uniformity of the intrachain exchange is expected since, regardless of faulting, each individual Cu-O-Cu connection within a chain is the same. 

The upturn and irreversibility observed in the magnetic susceptibility (Fig. \ref{fig:DC_magChar}(a)) indicate a deviation from quasi-1D behavior at $T <$ 75 K. One possibility is that this feature is associated with a spin glass transition. However, the lack of a clear, frequency-dependent peak in the AC magnetic susceptibility (Fig. \ref{fig:ACsuscept}) makes this unlikely. The temperature dependence of the broad peak observed in our neutron scattering data, which emerges around $T_{irr}$, is consistent with magnetic diffuse scattering. That this additional scattering occurs only at low 2$\theta$, with no additional temperature dependent scattering at high 2$\theta$, further indicates this scattering is magnetic and not due to a structural transition. As our heat capacity data (Figs. \ref{fig:heatCapacity_highT} and \ref{fig:heatCapacity_lowT}) shows no signatures consistent with long-range magnetic order we identify these features as due to the formation of short-range magnetic order which persists to low-temperature. 

%Although the magnetism at higher temperature displays a 1D character, the presence of extensive faulting of the Cu layers -- which is clearly demonstrated by our DIFFaX modeling (Fig. \ref{fig:refinement_diffax}) --  confirms the layered nature of the crystallographic structure. 

Based on the previous results it is clear that the disordered layering has a strong impact on the low-temperature magnetic properties of \CTOns. In an ideal 1D isotropic AFM Heisenberg model quantum fluctuations prevent the occurrence of long-range order at any temperature \cite{khomskii2010}. However, most real quasi-1D materials with isotropic Heisenberg exchange do eventually order due to the presence of finite interchain exchange interactions. The temperature scale of this order is largely determined by the ratio of the inter to intrachain interactions (see, for example, Table 4 in Ami \textit{et al.} \cite{ami1995}). 

The absence of long-range order down to temperatures significantly lower than the energy scale of the intrachain exchange interaction, despite the presence of short-range correlations at substantially higher temperatures, can be attributed to the high density of stacking faults. These stacking faults slightly modify the connectivity between Cu-chains in different layers  (see green and orange arrows in the insets of Fig. \ref{fig:refinement_diffax}). This in turn modifies the oribtal overlap within the Cu-O-Te-O-Cu pathway perpendicular to the Cu layers, which, we propose, creates a distribution of superexchange interactions in this direction. This distribution acts as a source of frustration preventing the occurence of long-range order across different Cu layers, and allowing only short-range order within the layers (at least within the temperature range explored here). 

Another possible interpretation is that the distribution of exchange interactions actually \textit{induces} short-range magnetic order into the system, which, in the absence of disorder, would show none. We are unable to definitively distinguish between these two interpretations due to the fact that the stacking faults are extended over an entire layer. This means that, while the disordered layering may not be the thermodynamically stable state, post-synthetic measures, such as annealing, are dynamically insufficient to resolve the structure to the P21/$n$ “ideal.” However, this interpretation seems unlikely in the context of the many other cuprates spin chains which are observed to order (\textit{e.g.} SrCuO$_{2}$ \cite{matsuda1997}, BaCu$_{2}$Si$_{2}$O$_{7}$ \cite{kenzelmann2001}, KCuF$_{3}$ \cite{hutchings1969}, and Cu(C$_{4}$H$_{4}$N$_{2}$)(NO$_{3}$)$_{2}$ \cite{lancaster2006}). Theoretical work by Yasuda \textit{et al.} \cite{yasuda2005} provides an empirical relation (equation 8 in their paper) for estimating the expected interchain coupling ($J'$) for Heisenberg spin chains based on the observed transition temperature ($T_{N}$) and $J_{1}$. Taking $T_{N}$ = 0.1 K (the lowest temperature for which we have data) as an upper bound, this relation gives $J'$/$J_{1}$ = 6.1(2) $\times$ 10$^{-4}$. This value would suggest that \CTO possesses some of the most isolated chains among quasi-1D cuprates \cite{lancaster2006,ami1995}, which appears unrealistic given the presence of clear Cu-O-Cu interchain exchange pathways within the Cu layers. 

This distinction aside, our physical picture is the formation of small, AFM, domains of coherently ordered chains \textit{within} a Cu layer (onsetting near $T_{irr}$) but with minimal coherence between distinct layers of chains. The correlation lengths from the Warren lineshape analysis at $T$ = 10 K and 40 K are on the order of the interlayer spacing (5.163(4) \r{A}), further reinforcing this picture.

Our neutron data are insufficient to definitively characterize the symmetry of this short-range magnetic order. However, the location of the diffuse magnetic scattering ($|\va*{Q}| \approx$ 0.7 \r{A}$^{-1}$) is close to the value for the (010) reflection ($|\va*{Q}|$ = 0.61 \r{A}$^{-1}$). This would indicate AFM order along the \textit{b}-axis, perpendicular to the Cu layers, with at least some component of the moment in the (101) plane. An alternative interpretation that we cannot definitively rule out is that this scattering reflects the appearance of magnetic order with a propogation vector of $\va*{k}$ = (0.5, 0, 0), with $|\va*{Q}|$ = 0.57 \r{A}$^{-1}$ for the first peak. However, this seems unlikely as the standard crystallographic cell of \CTO can accomodate in-plane magnetic order without enlarging the unit cell. While the precise symmetry of the short-range order remains an open question, the finite hysteresis in the magnetization curves  below $T_{irr}$ (Fig. \ref{fig:DC_magChar}(b)) indicates a weak FM component. This probably stems from a canting of the overall AFM order due to a finite Dzyaloshinskii-Moriya interaction, as has been seen in other layered cuprates \cite{bonesteel1993,coffey1991}. Another relevant point is that, due to the lack of energy discrimination in our diffraction measurement, we are unable to determine if there is an inelastic component to the observed diffuse scattering; future neutron spectroscopy measurements would be of interest to address this. We note that there is a slight feature in the magnetic susceptibility near $T \approx$ 20 K. The origin of this feature cannot be rigorously determined with the available data. However, we believe it is most likely due to the presence of a weak paramagnetic signal -- from either a small, unresolved impurity phase or defect spins within the \CTO matrix -- which is overlaid with the saturating signal from the weak ferromagnet component we have proposed. 

The linear term extracted  from the low-temperature heat capacity data is surprising given that \CTO appears to be a large band gap insulator, precluding the possibility of a high density of electronic states at the Fermi level. However, the uniform $S$ = $\frac{1}{2}$, AFM Heisenberg spin chain is known to possess a linear heat capacity at low temperature due to gapless spin excitations \cite{johnston2000,bonner1964,nagler1991} with the value of $\gamma$ given as $\gamma = \frac{2Rk_{B}}{3J_{1}}$ \cite{johnston2000}. Inputting our $J_{1}$ value yields $\gamma$ = 33(1) mJ mole$^{-1}$ K$^{-2}$, which is significantly higher than we observe ($\gamma$ = 9.58(8) mJ mole$^{-1}$ K$^{-2}$). This can be explained by the reduction of $J_{1}$ due to the increased ionicity of the Cu-O bond, as mentioned previously. It is also possible that the influence of interchain interactions, which perturb the system towards a higher dimensionality where no $T$-linear heat capacity term is expected, may also play a role. While our low-temperature heat capacity measurements allow us to qualitatively describe the magnetic contribution to the heat capacity, a quantitative determination is not currently possible due to the lack of a non-magnetic analog for \CTOns. However, at low temperatures the phonon contribution to the heat capacity often scales as the dimensionality of the material. The finite $T^{3}$ contribution ($\beta$ term) observed here therefore reflects the relatively 3D character of the phonon dispersion.
%\FloatBarrier
\section{Conclusion}

While \CTO was predicted to have a 2D AFM ground state \cite{botana2017}, extensive characterization enabled by our improved synthesis method demonstrates AFM Heisenberg spin chain behavior for $T \geq 75$ K with a uniform nearest-neighbor exchange of $J_{1}$ = 164(5) K. Careful conisderation of the Cu connectivity indicates this behavior is due to the arrays of corner connected CuO$_{4}$ plaquettes highlighted in Fig. \ref{fig:StructureComparison}(b,c). Although no long-range order was detected in our heat capacity measurements, the combination of an upturn in the magnetic susceptibility and diffuse scattering in the neutron diffraction data indicate persistent short-range magnetic order below $T$ = 60 K. The persistence of these short-range correlations down to temperatures well below the energy scale of the intrachain exchange interaction is attributed to the presence of a significant fraction of stacking faults. These faults modify the interlayer connectivity and creates a distribution of interchain interactions that frustrates any incipient magnetic order. The low-temperature heat capacity shows an unexpectedly large linear term for a wide band gap insulator which is likely related to the quasi-1D magnetism. 
 
 Although the first principles calculations in Botana \textit{et al.} \cite{botana2017} showed \CTO to have a small bandgap of 0.13 eV, making it a suitable candidate for superconductivity upon doping, this appears inconsistent with the yellow color of the polycrystalline powder. The underestimation of the band gap may be due to the failure of DFT to consider the effect of hypervalent cations like Te$^{6+}$ which has a strong inductive withdrawal effect on O$^{2-}$. Additionally, we find no evidence of dopant solubility with our synthesis method. These findings suggest that future efforts towards discovering high-temperature superconductivity in copper tellurates should be focused on cases where there is a reduced inductive withdrawal effect by Te, such as those compounds where Te is present in the 4+ state.

\begin{acknowledgments}
This work was supported as part of the Institute for Quantum Matter, an Energy Frontier Research Center funded by the U.S. Department of Energy, Office of Science, Office of Basic Energy Sciences, under Award No. DE-SC0019331. The MPMS was funded by the National Science Foundation, USA, Division of Materials Research, Major Research Instrumentation Program, under Award No 1828490. K.E.A acknowledges support from the Humboldt Foundation Postdoctoral Fellowship. T.T.T acknowledges support from Clemson University, College of Science, Department of Chemistry. E.Z. acknowledges support from the Sweeney Family Postdoctoral Fellowship.
\end{acknowledgments}

% Create the reference section using BibTeX:
%\bibliography{Bibliography}

%apsrev4-2.bst 2019-01-14 (MD) hand-edited version of apsrev4-1.bst
%Control: key (0)
%Control: author (8) initials jnrlst
%Control: editor formatted (1) identically to author
%Control: production of article title (0) allowed
%Control: page (0) single
%Control: year (1) truncated
%Control: production of eprint (0) enabled
%

\end{document}